\documentclass[twocolumn,10pt,a4paper,sort&compress,aps,pre,showpacs,superscriptaddress]{revtex4-2}
\usepackage[english]{babel}
\usepackage[caption=false]{subfig}
\addto\captionsenglish{}

\usepackage{overpic}
\usepackage{amsmath,amssymb,amsthm}
\usepackage{commath}
\usepackage{graphicx}
\usepackage{dcolumn}
\usepackage{bm}
\usepackage{hyperref}
\usepackage{etoolbox} 
\usepackage{siunitx}
\usepackage{multirow}

\usepackage[capitalise,nameinlink]{cleveref}
\newcommand{\aref}[1]{\hyperref[#1]{Appendix}}
\usepackage{blkarray}

\usepackage[normalem]{ulem}
\usepackage{comment}
\usepackage[]{xcolor}

\begin{document}

\title{Robust undulatory locomotion via neuromechanical adjustments\\ in a dissipative medium}

\author{Kenta Ishimoto}
\email{ishimoto@kurims.kyoto-u.ac.jp}
\affiliation{Research Institute for Mathematical Sciences, Kyoto University, Kyoto 606-8502, Japan}

\author{Cl\'{e}ment Moreau}
\email{clement.moreau@cnrs.fr}
\affiliation{Nantes Universit\'e, \'Ecole Centrale Nantes, CNRS, LS2N, UMR 6004, F-44000 Nantes, France}

\author{Johann Herault}
\email{johann.herault@imt-atlantique.fr}
\affiliation{LS2N, IMT Atlantique, 44307 Nantes, France}

\date{\today}

\begin{abstract}
Dissipative environments are ubiquitous in nature, from microscopic swimmers in low-Reynolds-number fluids to macroscopic animals in frictional media. In this study, motivated by various behaviours of {\it Caenorhabditis elegans} during swimming and crawling locomotion, we consider a mathematical model of a slender elastic locomotor with an internal rhythmic neural pattern generator. By analysing the dynamical systems of the model using a Poincar\'e section, we found that local neuromechanical adjustments to the environment can create robust undulatory locomotion. This progressive behaviour emerges as a global stable periodic orbit in a broad range of parameter regions. Further, by controlling the mechanosensation, we were able to
design the dynamical systems to manoeuvre with progressive, reverse, and turning motions as well as apparently random, complex behaviours, as experimentally observed in {\it C. elegans}. The mechanisms found in this study, together with our methodologies with the dynamical systems viewpoint, are useful for deciphering complex animal adaptive behaviours and also designing adaptive robots for a wide range of dissipative environments. 
\end{abstract}

\maketitle

\section{Introduction}

Biolocomotion is very diverse, and its gait pattern varies in different media, including swimming in water, gliding and flying in air, and walking and crawling on the ground. 
Undulatory locomotion for swimming and crawling is used by various species, from flagellated cellular locomotion in invertebrates to swimming and crawling by vertebrates, such as fish and snakes \cite{nosrati2015two, gazzola2014scaling, jayne2020defines, di2021convergence, guasto2020flagellar, gaffney2021modelling}.

It has long been a major challenge to find the common biological mechanisms that enable adaptation to different mechanical environments.
In some vertebrates, such as lamprey, central pattern generators (CPGs) in the nervous system have been well known as the key mechanism underlying gait control and environmental adaptation. CPGs are made of networks of neural oscillators in the spinal cord, where autonomous neuronal circuits produce cyclic motor activity. This rhythmic pattern generates  basic locomotion and can be modulated by sensory feedback to adapt to the environment \cite{ijspeert2008central, wen2018caenorhabditis,thandiackal2021emergence, tsybina2022toward, wyart2023design}.

Animal mechanosensation has been intensively studied in fish undulatory swimming at moderate and large Reynolds numbers, with a focus on proprioceptive feedback, which uses perceptions of local body deformation to sustain a desired gait pattern via stretch feedback \cite{tytell2010interactions, hamlet2023proprioceptive, gordleeva2023control}. Similarly, these aquatic animals can also perceive the dynamics of the fluid, giving rise to exteroceptive sensory feedback, through their lateral line or mechanical receptors \cite{buchanan1982activities, di2000cellular, ristroph2015lateral}. The role of exteroceptive feedback in local sensorimotor loops is an ongoing research subject, but recent experiments with fish-like swimming robots in water  \cite{thandiackal2021emergence}  found that a robust undulatory locomotion can be achieved by local exteroceptive feedback control. 
 

\begin{figure}[t]
\begin{center}
\begin{overpic}[width=8.5cm]{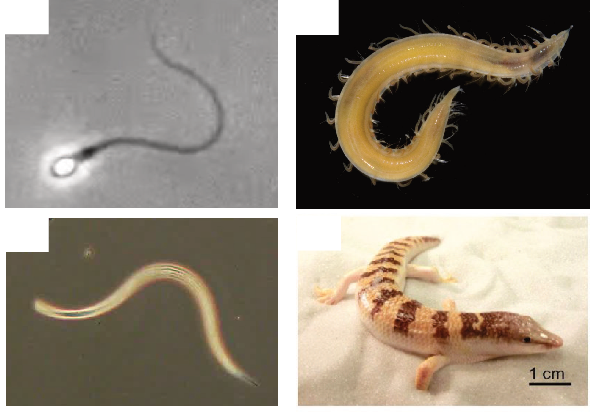}
\put(1.5,66){(a)}
\put(50.5,66){(b)}
\put(1.5,29){(c)}
\put(50.5,29){(d)}
\end{overpic}
\caption{ Examples of undulatory locomotion in a strongly dissipative medium. (a) Human sperm cells having whip-like motion without a neural network. Videomicroscopy from Movie 1 of the Supplemental Material of  \cite{ishimoto2017coarse}.
(b) Polychaeta worm, {\it Armandia brevis}, that  can burrow in mud. Image adapted from \cite{wbst} under CC BY-SA license (copyright 2024, Florida Museum of Natural History). 
(c) Swimming nematode captured and recorded by the authors at the north campus of Kyoto University. (d) Sandfish lizard that can swim in sand. Image adapted from \cite{ding2013emergence} (copyright 2013, National Academy of Sciences).}
\label{fig:images}
\end{center}
\vspace*{-0.5cm}
\end{figure}
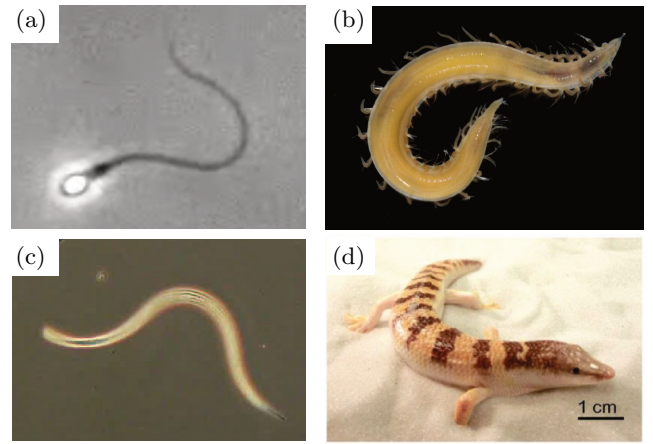

These vertebrates live in a large-Reynolds-number regime,  where the inertia of the body and fluid plays a crucial role. However, undulatory locomotion is used even when the inertia is negligible, and this regime is the focus of this study.

Animal locomotion in dissipative media includes swimming in low-Reynolds-number flow and crawling locomotion on a frictional surface or granular medium \cite{ishimoto2017coarse, hosoi2015beneath, rieser2019geometric} (Fig. \ref{fig:images}). In such a dissipation-dominated environment, the drag is proportional to the body velocity, and for an elongated body, empirical resistive force theory (RFT), which only considers local tangential and perpendicular drag coefficients, is known to be widely applicable across all scales in a highly dissipative system \cite{friedrich2010high, schulman2014dynamic, keaveny2017predicting, hosoi2015beneath, rieser2019geometric}, including sandfish lizards swimming in sand \cite{ding2013emergence}, polychaeta worms burrowing in mud \cite{dorgan2013meandering} and snakes crawling on sand \cite{rieser2019geometric}.

In the highly dissipative and inertialess regime, the kinematic reversibility of the system necessitates non-reciprocal deformation of a self-propelled object to achieve net locomotion, which is well known as the scallop theorem \cite{purcell1977life, ishimoto2012coordinate} in low-Reynolds-number hydrodynamics. For instance, a complete in-phase synchronisation of motor activation of ipsilateral muscles leads to a reciprocal deformation; hence, the animal should break the reciprocity of the internal states. In contrast, some fish can still swim in this case, a gait known as oscillatory swimming  at large Reynolds numbers. Consequently, non-reciprocity in highly dissipative  media imposes specific constraints on motor activity.

Neuromechanics in this regime have been studied intensively for a nematode, {\it Caenorhabditis elegans}, a model organism in neuroscience. Recent studies suggest the existence of a CPG-like network in {\it C. elegans}, reporting multiple sections of rhythmic cycles \cite{fouad2018distributed}, and a relaxation-oscillator model supports these experimental findings \cite{ji2021phase}. 

The rhythmic neural activity of {\it C. elegans} has been used to develop integrated neuromechanical models for its undulatory locomotion \cite{boyle2012gait, cohen2019whole}. More recently, based on RFT, fluid-structure interactions have been incorporated into integrated models that contain muscle activation and neural coupling with proprioceptive feedback. These models have successfully reproduced gait transitions of {\it C. elegans} between swimming and crawling and waveform modulation for swimming in media with different viscosities \cite{denham2018signatures,johnson2021neuromechanical}. 
Johnson et al. \cite{johnson2021neuromechanical} further theoretically derived a coarse-grained model from their integrated model by phase-oscillator reduction and found that the reduced model well captures the gait adaptation of swimming in {\it C. elegans}, motivating a simplified and minimal mathematical model for animal gait adaptation in dissipative media. 

Hence, in this study, we developed a simple and universal algorithm for neuromechanical adaption to different environments in a dissipation-dominated regime, by using local exteroceptive mechanical feedback from the environment. To do so, we extend the coupled-oscillator model with fluid-structure interactions used in \cite{thandiackal2021emergence}, which reported robust self-organised locomotion via local hydrodynamic sensing in the inertial regime.


Our primary aim is therefore to understand the mechanisms of neuromechanical adjustment and complexity of animal behaviour in a dissipation-dominated regime through a simple and minimal mathematical model. Therefore, we introduce a high-level model of motor activation taking the form of CPGs that produce reciprocal motion in the absence of external stimuli.
The underlying motivation is to introduce a fundamental motor control  that does not favour any particular propulsive gait, due to motion reciprocity. By introducing sensory feedback in this specific CPG, we show that locomotion can be an emerging property of couplings between the body dynamics and the neuromotor system in dissipative media. Indeed, we report various motions, such as crawling and swimming at low and high viscosity, resulting from the adaptation to mechanical properties of the swimmer or drag of the dissipative medium.

To describe the diversity of the observed motions, we took inspiration from the behavioural-state space representation of {\it C. elegans}, recently introduced in an experimental study \cite{ahamed2021capturing}. In this framework, its behaviour can be fully characterised by the state of its mechanical system and body shape. Our second aim is to extend this approach by considering the full neuromechanical state of the animal and to analyse and design complex behaviours in terms of smooth trajectories produced by our neuromechanical model.

To reproduce behavioural variability, we hypothesised that this smooth trajectory is ruled by a non-autonomous dynamical system displaying different attractors. Here, the gait is represented by a periodic orbit or limit cycle, while the animal posture corresponds to a fixed point. We show how couplings between body-environment dynamics and CPGs can generate transient dynamics between stable and unstable periodic orbits, which can reproduce reversal of the motion direction. Motivated by the turning motion of {\it C. elegans}, we also introduced a phase-locking mechanism in the CPG to create a local fixed point in the state space corresponding to a posture producing the omega-turn manoeuvre \cite{stephens2008dimensionality}. Finally, we demonstrated how alternating sequences of the phase-locked regime and sensory-driven locomotion can mimic foraging strategies in {\it C. elegans}. 

The content of the paper is as follows. Sec. \ref{sec:model} introduces our mathematical model of a slender elastic object in a dissipative medium. The inner muscle activity is driven by the rhythmic pattern generated by CPGs, and the CPG phase is described by a coupled oscillator, which is locally modulated via mechanosensory feedback. We present typical emergent behaviours in Sec. \ref{sec:sims} and further analyse the dynamical system with a focus on the trajectories in a Poincar\'e section in Sec. \ref{sec:stab}. In Sec. \ref{sec:turn}, by controlling the phase dynamics, we evaluate a manoeuvring algorithm that reproduces omega-turns of {\it C.elegans}. Concluding remarks are made in Sec. \ref{sec:conc}.

\section{Model}
\label{sec:model}

\subsection{Body-environment coupling}

We modelled the animal body as an inextensible elastic slender rod of length $L$. The centreline is denoted by its arclength $s\in [0, L]$, and its position by $\bm{x}(s,t)$. To calculate the force and torque balance for the local rod segment, contributions from elasticity, drag from the environment and inner activation were incorporated.

\begin{figure}[!t]
\begin{center}
\begin{overpic}[width=8cm]{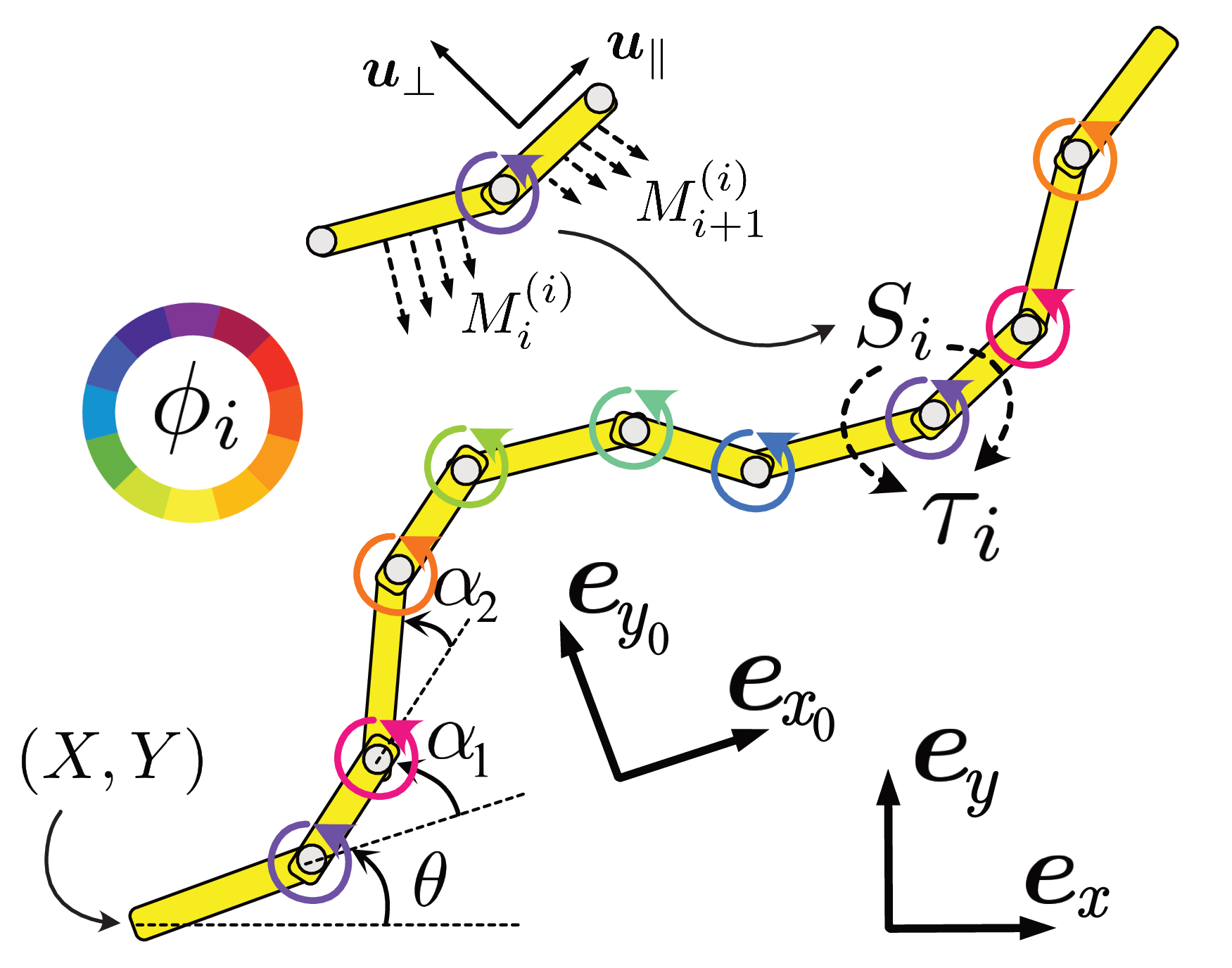}
\end{overpic}
\caption{Schematic of the model and the two frames of reference. A slender elastic rod is represented by $N$ links connected by elastic hinges. Each hinge generates internal torque $\tau_i$ $(i=1,2,\dots,N-1)$, as a function of the phase $\phi_i(t)$, which models the local central pattern generators. The phase dynamics are described by a coupled oscillator equation with modulation by local signal $S_i(t)=M_i^{(i)}+M_{i+1}^{(i)}$, with $M_i^{(j)}$ being the  drag on the $i$-th link around the $j$-th hinge (inset). The laboratory-fixed frame $\{ \bm{e}_x, \bm{e}_y, \bm{e}_z\}$ and body-fixed frame $\{ \bm{e}_{x_0}, \bm{e}_{y_0}, \bm{e}_{z_0}\}$ are introduced, and the motions are assumed to be restricted in the $x-y$ plane. The motion of the object is represented by the end of the first link $(X, Y)$ and its angle to the laboratory frame $\theta$. The shape of the body is described by $N-1$ relative angles $\alpha_i$.
}
\label{fig:config}
\end{center}
\end{figure}

We assumed planar locomotion (we set this as our $xy$ plane) and no twisting motion. 
In biological locomotors, muscular activation usually occurs through actuation of several elements along the body. 
For instance, {\it C. elegans} has approximately six muscular modules on its body \cite{johnson2021neuromechanical}.
This discrete activation structure within a continuous elastic rod is conveniently described by the coarse-grained representation of the elastic slender rod, known as the $N$-link model, where the object is represented by $N$ equal-length links connected at $N-1$ hinges \cite{moreau2018asymptotic, walker2019filament} (Fig. \ref{fig:config}). 
Here, we take $N=10$, following a previous study \cite{thandiackal2021emergence}.
We write the position of the end of the first link as $(X, Y)$ and the angle from the $x$-axis as $\theta$. As shown in Fig. \ref{fig:config}, these represent the position and orientation of the body and define the body-fixed frame $\{\bm{e}_{x_0},\bm{e}_{y_0},\bm{e}_{z_0}\}$, which is distinct from the laboratory frame $\{\bm{e}_x,\bm{e}_y,\bm{e}_z\}$. The object shape is described by $N-1$ relative angles between the neighbouring links and denoted by $\alpha_i$ ($i=1,2,\dots, N-1)$.

For the body-environment coupling, we used the resistive force theory, in which the local force per unit length at $\bm{x}$ is assumed to be proportional to the local tangent and normal velocities, $\bm{u}_{\parallel}$ and $\bm{u}_{\perp}$, as
$\bm{f}^{\textrm{env}}=-c_{\parallel}\bm{u}_{\parallel}-c_{\perp}\bm{u}_{\perp}$, where the coefficients, $c_{\parallel}$ and $c_{\perp}$, are the drag coefficient in each direction (Fig. \ref{fig:config} inset).
Due to the linearity of this body-environment coupling, 
the force and torque balance equations for each body segment 
can be summarised in matrix form \cite{ishimoto2022self, ishimoto2023odd}:
\begin{equation}
{\bf A}\dot{\bm{z}}=\bm{t}
    \label{eq:M03}
\end{equation}
Here, $\dot{\bm{z}}$ is a generalised velocity of the moving and deforming body as $\dot{\bm{z}}=(\dot{X}_0, \dot{Y}_0, \dot{\theta}, \dot{\alpha}_1, \cdots, \dot{\alpha}_{N-1})^{\textrm{T}}$, where the dot symbol indicates the time derivative and $(\dot{X}_0, \dot{Y}_0)$ is the velocity vector in the body-fixed frame. The matrix ${\bf A}(\bm{\alpha})$ is the generalised resistance matrix \cite{doi2013soft}, which encodes the mechanical interactions with the surrounding environment and is a function of the instantaneous shape. The vector $\bm{t}$ on the right-hand side is the generalised force and torque vector, and for free locomotion only the bottom $N-1$ rows may have non-zero values as $\bm{t}=(0, 0, 0, \tau_1, \dots, \tau_{N-1})^{\textrm{T}}$. Here, the torque at the $i$-th hinge 
contains elastic bending and internal actuation, which we denote $\tau_i=\tau_i^{\textrm{ela}}+\tau_i^{\textrm{int}}$.

\subsection{Coupled oscillators and mechanosensation}

We assume that the inner activation is controlled by a network of interconnected neural oscillators forming the CPGs, which provide rhythmic motions to the junctions. To model the cyclic activation of the junctions, we assume that the $i$-th oscillator is associated with the corresponding hinge, and generates a periodic activation signal given by a phase denoted as $\phi_i(t)$ $(i=1,2,\dots,N-1)$ (Fig. \ref{fig:config}). 

For modelling the elasticity of the body, we employ the Euler--Bernoulli constitutive relation, which assumes a linear relation between the bending moment and local curvature. With a bending modulus $\kappa$ for the discretised rod, we then have $\tau_i^{\textrm{ela}}=-\kappa\alpha_i$. For the torque actuation, we simply assume a periodic sinusoidal mode \cite{thandiackal2021emergence}, given by $\tau_i^{\textrm{int}}=\tau\cos\phi_i(t)$, where
$\tau$ is the torque actuation strength at each hinge. The torque function is then summarised as
$\tau_i=-\kappa \alpha_i+\tau\cos\phi_i(t)$.

Following previous studies of CPGs for undulatory locomotion \cite{cohen1992modelling, ijspeert2007swimming} to describe this system, we describe the oscillator dynamics by a limit cycle close to a Hopf bifurcation.  In this case, the phase reduction of the dynamics leads to \cite{Kuramoto} 
\begin{equation}
\dot{\phi_i}=\omega_0+\sum_j C_{ij}\sin(\phi_j-\phi_i)+Z(\phi_i)S_i(t)
    \label{eq:M02},
\end{equation}
where we consider a symmetric, nearest-neighbour phase coupling with $C_{i,i\pm1}=C$ and zero otherwise. 
The phase oscillators are assumed to have a common intrinsic frequency $\omega_0$ and coupling with their neighbours with a common strength that is modulated by sensory feedback from the surrounding mechanical environment. 
Close to a Hopf bifurcation, the phase sensitivity function takes the form of $Z(\phi_i)=\sigma \cos\phi_i(t)$ with a sensitivity strength $\sigma$   \cite{thandiackal2021emergence}. 

The phase sensitivity function $Z(\phi_i)$ characterises the response of a limit cycle weakly perturbed by the sensory feedback $S_i(t)$. The function $Z(\phi_i)$ modulates the input signal $S_i(t)$ such that their correlation modifies the phase dynamics, resulting in phase acceleration or deceleration. 
Synchronisation and phase locking between these two signals is an emergent phenomenon that depends on both the choice of the sensitivity function and the nature of the sensory feedback. 

As the sensory feedback, we employed the local torque load at each hinge, following the model of swimming lamprey with exteroception \cite{thandiackal2021emergence} and rhythmic contraction of true slime mould \cite{kobayashi2006mathematical, kobayashi2013design}. 
Let $M^{(j)}_{i}$ be torque (per unit length) from the environmental drag on the $i$-th link around $j$-th hinge, or equivalently,
\begin{eqnarray}
    M^{(j)}_i=\bm{e}_z  \cdot \int_{s_{i-1}}^{s_{i}} \left ( \bm{x}(s) - \bm{x}(s_j) \right ) \times \bm{f}^{\mathrm{env}}(s)\,\frac{\mathrm{d} s}{\Delta s},
\end{eqnarray}
where  $\Delta s=L/N$ and $s_i=i \Delta s$ corresponds to the position of the $i$-th hinge (inset of Fig.\ref{fig:config}). 
We then set a fore-aft symmetric feedback $S_i(t)=M^{(i)}_i+M^{(i)}_{i+1}$, which is the local torque load on the $i$-th hinge. 

Other choices to model proprioceptive sensory feedback include employing the local curvature as signal input, which, however, must be asymmetric to break the fore-aft symmetry and generate undulatory locomotion. 
Some mathematical models consider asymmetric and non-local signal inputs with weight on the posterior neighbour (e.g., \cite{johnson2021neuromechanical}), enforcing a directed motion. 
In contrast, our exteroceptive feedback is a local and symmetric function, which fits our purpose to simplify the mechanism for adaptive locomotion.  

\subsection{Model parameters}

The mechanical torque on the body is calculated through the resistive force theory, which is applied for various undulatory motions in dissipation-dominated environments, such as low-Reynolds-number flows, gel-like structures, and mud and sand. 
We then introduce the anisotropic  drag ratio as $\gamma=c_{\parallel}/c_{\perp}$, which is known to affect locomotion. Following the data obtained for {\it C. elegans}, we use $\gamma=1/2$ for {\it swimming} and $\gamma=1/70$ for {\it crawling} \cite{schulman2014dynamic, keaveny2017predicting}. When the model swims in a low-Reynolds-number Newtonian fluid, the coefficients are theoretically estimated \cite{lighthill1976flagellar} as $c_{\perp}=4\pi\mu/[\log(2L/d)]$ in a slender asymptote, where $\mu$ is the fluid viscosity and $d$ is the cross-sectional radius of a slender body.

For our numerical and theoretical analyses, we non-dimensionalised the above system by setting the scales $L=1$ for length, $T=2\pi/\omega_0=1$ for time, and $\kappa=1$ for force.
An important non-dimensional number of the system is the so-called `sperm number', which represents the effective flexibility of an oscillatory elastic rod in a highly damped system, defined as $\textrm{Sp}=(\omega_0 c_{\perp}/\kappa L^3)^{1/4}$ \cite{moreau2018asymptotic}.

A typical size of $\textrm{Sp}$ for {\it C. elegans} swimming in water may be obtained as $\textrm{Sp}\approx 4.2$, if we choose $L=10^{-3}$ m, $\mu=10^{-3}$ Pa$\cdot$s and the experimental data of \cite{fang2010biomechanical}, $\kappa_b= \kappa L N =9.5\times 10^{-13}$ N$\cdot$m$^2$ and $1/T=1.7$ Hz, although other studies reported smaller values of bending modulus $\kappa_b$  \cite{johnson2021neuromechanical}.

 Other parameters are the drag coefficient ratio $\gamma$, non-dimensionalised actuation strength $\hat{\tau}$, coupling strength $\hat{C}$ and sensitivity strength $\hat{\sigma}$. 
To break reciprocity in the CPG, we progressively increase the weight of sensory feedback in oscillator dynamics by exploring a wide range of parameters $\hat{\sigma}/\hat{C}$ ranging from $10^{-2}$ to $10^2$. 
Hereafter, we omit the hat symbol for the non-dimensionalised strengths.


\section{Robust undulatory locomotion}
\label{sec:sims}

\begin{figure}[!t]
\begin{center}
\begin{overpic}[width=8cm]{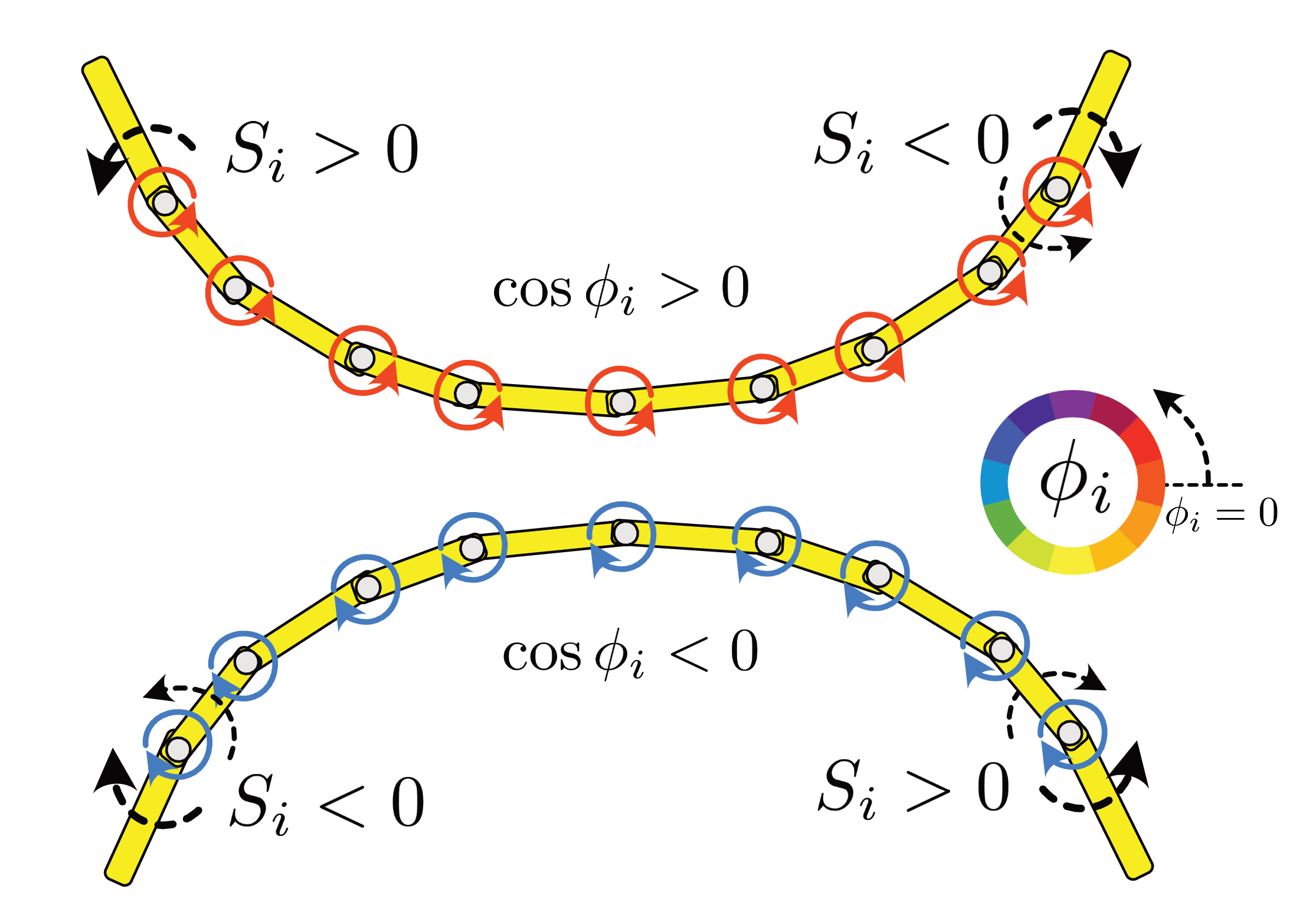}
\put(-1,64){(a)}
\put(-1,28){(b)}
\end{overpic}
\caption{Schematic of the mechanism of symmetry breaking via mechanosensory feedback. Fully synchronised situations are illustrated with (a) $\cos\phi_i>0$ and (b) $\cos\phi_i<0$. The direction of the local torque actuation is opposite, and the shape becomes concave in (a) and convex in (b). Despite the uniform distribution of the torque, the local torque load is no more uniform along the object, yielding fore-aft asymmetric phase modulation due to mechanoreception. In both phases, the product of $S_i$ and $\cos\phi_i$ is always positive at the left end and negative at the right end. Thus, when the sensitivity strength, $\sigma$, is positive, the phases are accelerated at the left and decelerated at the right, leading to a travelling undulatory motion. }
\label{fig:sym}
\end{center}
\end{figure}

\begin{figure*}[!t]
\begin{center}
\begin{overpic}[width=6cm]{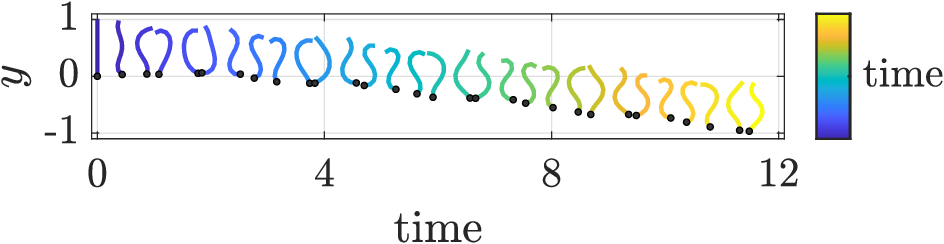}
\put(-1,27){(a)}
\put(18,28){Swimming (low viscosity)}
\put(75,20){$\downarrow$}
\end{overpic}
\begin{overpic}[width=6cm]{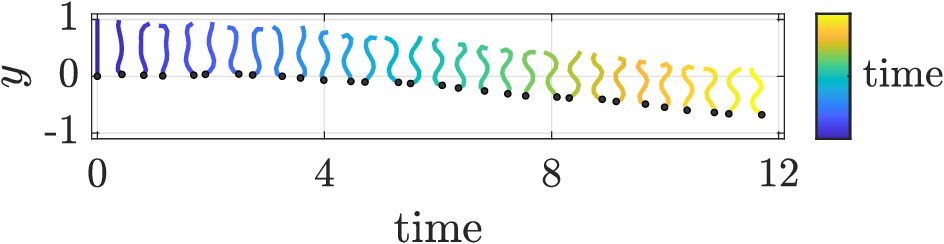}
\put(-1,27){(b)}
\put(16,28){Swimming (high viscosity)}
\put(79,20){$\downarrow$}
\end{overpic}
\begin{overpic}[width=5cm]{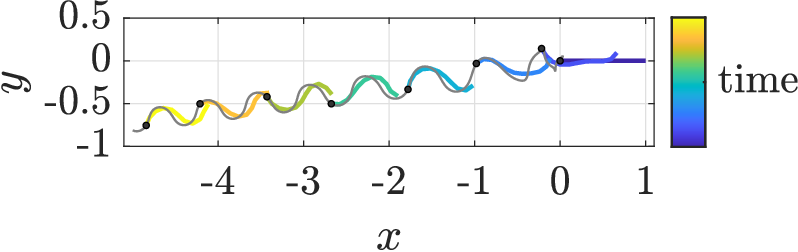}
\put(0,32){(c)}
\put(35,33){Crawling}
\put(52,24){$\longleftarrow$}
\end{overpic}\\
\begin{overpic}[width=6cm]{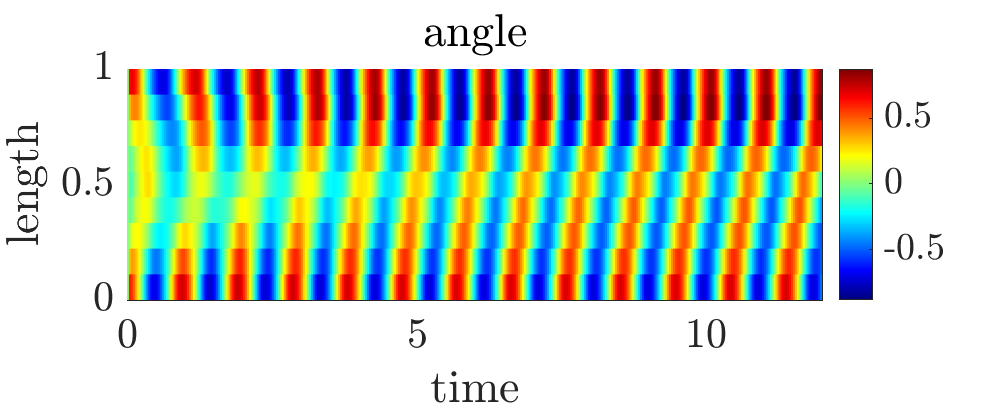}
\end{overpic}
\begin{overpic}[width=6cm]{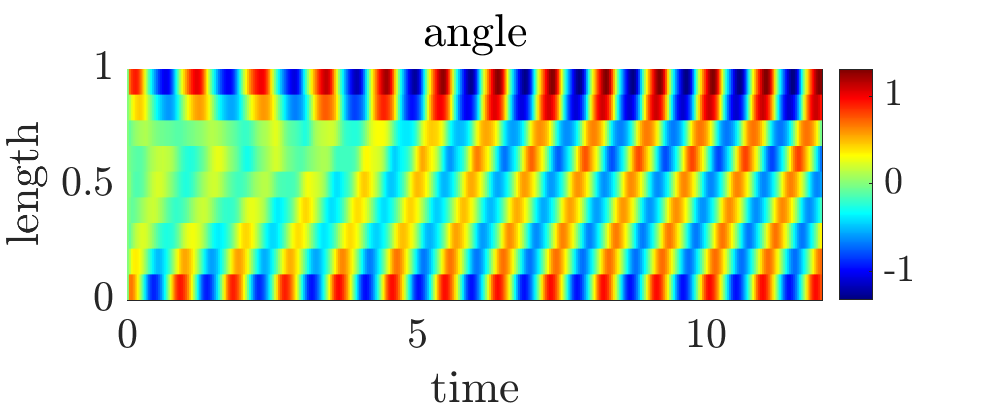}
\end{overpic}
\begin{overpic}[width=5cm]{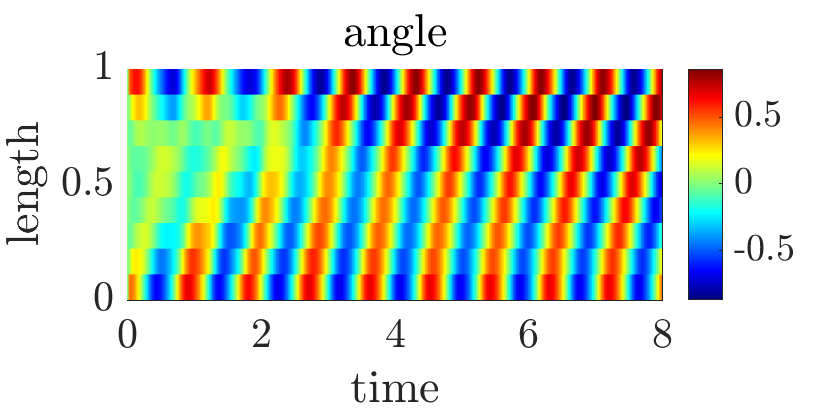}
\end{overpic}\\
\begin{overpic}[width=6cm]{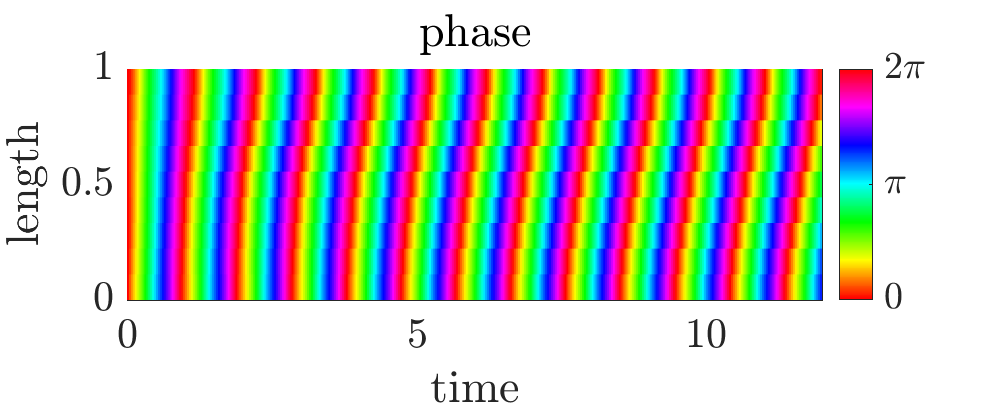}
\end{overpic}
\begin{overpic}[width=6cm]{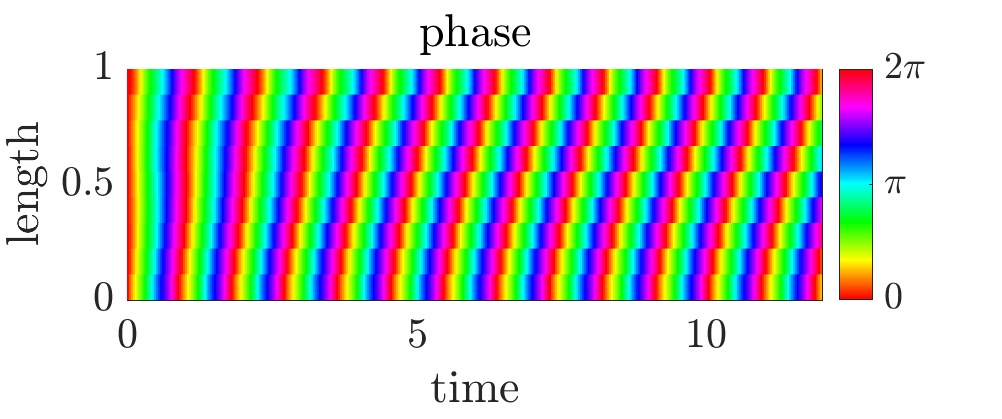}
\end{overpic}
\begin{overpic}[width=5cm]{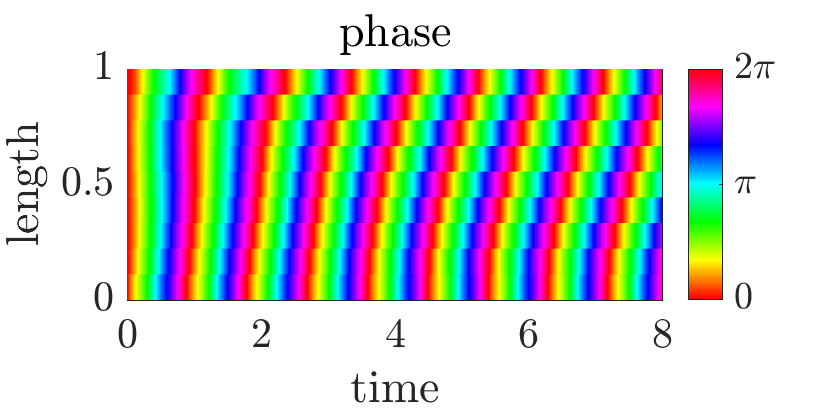}
\end{overpic}\\
\caption{Typical emergent motions of (a) swimming in a low-viscosity medium, (b) swimming in a high-viscosity medium, and (c) crawling on an agar gel. The model animal was initially aligned in a straight configuration with the same phase. The top panels of (a) and (b) show the time series of shape gaits. The top panel of (c) presents snapshots of the shape in the $xy$ plane superimposed on the trajectory. The shape of the locomotor is plotted in different colours to present the time evolution and the black dot denotes the end of the first link in each snapshot. The time series of the angle $\alpha_i$ and the phase $\phi_i$ are shown as kymographs in the middle and bottom panels.} 
\label{fig:sims}
\end{center}
\end{figure*}


Before proceeding to the simulation results, we first consider how motion non-reciprocity can be produced by breaking the symmetry with respect to the locomotor's transverse plane in the oscillator chain from the fore-aft symmetric sensory feedback function. The schematic in Fig. \ref{fig:sym} explains the symmetry-breaking mechanism. In Fig. \ref{fig:sym}(a), the oscillators are fully synchronised with $\cos\phi_i>0$ such that the counterclockwise torque generates a concave shape, where the end links rotate more quickly. The local torque load is therefore counterclockwise at the left end $(S_i>0)$, while the torque load is clockwise at the right end $(S_i<0)$, yielding a negative gradient of resistive torque along the rod. This gradient is then projected along the oscillator through the correlation $\cos(\phi)S_i$ in the phase equation [Eq. \ref{eq:M02}].

When the synchronised phases satisfy $\cos\phi_i<0$ [Fig. \ref{fig:sym}(b)], the direction of the local torque actuation reverses and the shape becomes convex. As in Fig. \ref{fig:sym}(b), the local torque load has the opposite sign in the left and right ends but with the opposite sign of $S_i$ for the case of Fig. \ref{fig:sym}(a), that is, a positive gradient of $S_i$. Nonetheless, the modulation to the phase dynamics, a product of $S_i$ and $\cos\phi_i$, remains positive at the left end and negative at the right end in both phases. Thus, when the sensitivity strength, $\sigma$, is positive (negative), the phases are accelerated on the left (right) side of the object, and decelerated on the right (left) side of the object, leading to symmetry breaking at the origin of the travelling wave of motor activation.


We then evaluated the simulation results in a different medium by changing the sperm number $\textrm{Sp}$ and drag anisotropic ratio $\gamma$. In a wide parameter regime, we found that the object eventually exhibits stable undulatory locomotion by breaking the fore-aft symmetry (Fig. \ref{fig:sym}). The direction of motion is governed by the sensitivity strength sign, and we found forward motion (moving towards the first link) when $\sigma>0$ and backward motion (towards the last link) when $\sigma<0$. 

In Fig. \ref{fig:sims}(a), we show typical dynamics of swimming motion in a watery low-viscosity fluid with $\textrm{Sp}=4$ and $\gamma=1/2$. The top panel shows snapshots of the shape of the model in different colours.
The model animal, initially aligned in a straight configuration, started to create a travelling wave and eventually exhibited stable periodic swimming. The middle and bottom panels present the kymographs of the angle $\alpha_i(t)$ and the phase $\phi_i(t)$. The other parameters were set as $\tau=8$, $\sigma=4$ and $C=1$. 

In Fig. \ref{fig:sims}(b), we show typical dynamics of swimming motion in a more viscous fluid with $\textrm{Sp}=6$ and $\gamma=1/2$.
Again, we found that undulatory locomotion eventually emerges after some transient phase. 
We confirmed that increasing the viscosity ($\mathrm{Sp}$) leads to a shorter wavelength and this tendency is very robust along a broad range of model parameters. The values of $C$ and $\sigma$ in the coupled oscillator equation mostly contribute to the transient timescale for reaching periodic locomotion.
The parameters used in the figure are $\tau=12$, $\sigma=8$ and $C=1$. A larger torque generation was required to sustain the wave amplitude, which implies that a stronger muscle contraction is generated in a high-viscosity medium, with this being in agreement with previously known experimental observations of {\it C. elegans} \cite{fang2010biomechanical}.

We then studied crawling behaviour by setting the drag anisotropy ratio as $\gamma=1/70$ \cite{keaveny2017predicting} while keeping the sperm number $\textrm{Sp}=6$.
Decreasing $\gamma$ leads to a larger undulatory motion wavelength in crawling. More importantly, the locomotion trajectory begins to follow the body shape, which is seen in {\it C. elegans} \cite{pierce2008genetic}. In the top panel of Fig. \ref{fig:sims}(c), we plot superimposed snapshots in the physical $x-y$ space to show the actual trajectory and shape.
The parameter set used in the figure was $\tau=8$, $\sigma=12$ and $C=0.1$.

In this section, we have seen that the local mechanosensory feedback based on the local torque load breaks the fore-aft symmetry and generates a robust undulatory locomotion in a self-organised manner. The exteroceptive feedback control is therefore a simple but prominent mechanism to adapt to different mechanical environments even in dissipative media. 
\section{Stability of periodic orbits}
\label{sec:stab}

In the previous section, stable, robust periodic locomotion was observed in a wide range of parameters. 
We now further analyse the underlying dynamical system to characterise the existence and stability of these emerging periodic orbits. 


We introduce the dynamics state for a generic $N$-link model as 
\begin{equation}
\bm{x}=(\alpha_1, \alpha_2,\dots, \alpha_{N-1},\phi_1,\phi_2,\dots\phi_{N-1} )^{\textrm{T}}\in U,
\end{equation}
where $U=\mathbb{R}^{N-1}\times\mathbb{T}^{N-1}$ is the state space. We choose our Poincar\'e section at $\phi_1=0$ and denote the section as $P=\mathbb{R}^{N-1}\times\mathbb{T}^{N-2}$, which leads to the associated Poincar\'e map $F:P\rightarrow P$ [Fig. \ref{fig:purcell}(b)].
Here, we consider a periodic orbit in the state space including its configuration and CPG phases, while the CPG phase is a coarse-grained representation of the oscillatory neural dynamics at each junction.

To detect periodic orbits in $U$, which is regarded as a fixed point in $P$ such that $F(\bm{x}^\ast)=\bm{x}^\ast$, we use the standard Newton--Raphson method \cite{ishimoto2014study} via $\bm{x}_{m+1}=\bm{x}_m-[{\bf J}(\bm{x}_m)]^{-1}[F(\bm{x}_m)-\bm{x}_m]$ for an integer $m$, where ${\bf J}(\bm{x}_m)$ represents the Jacobian matrix. The eigenvalues of the Jacobian matrix approximate the Floquet exponents when the algorithm converges.

\subsection{Purcell's three-link model}

\begin{figure}[!t]
\begin{center}
\begin{overpic}[width=9cm]{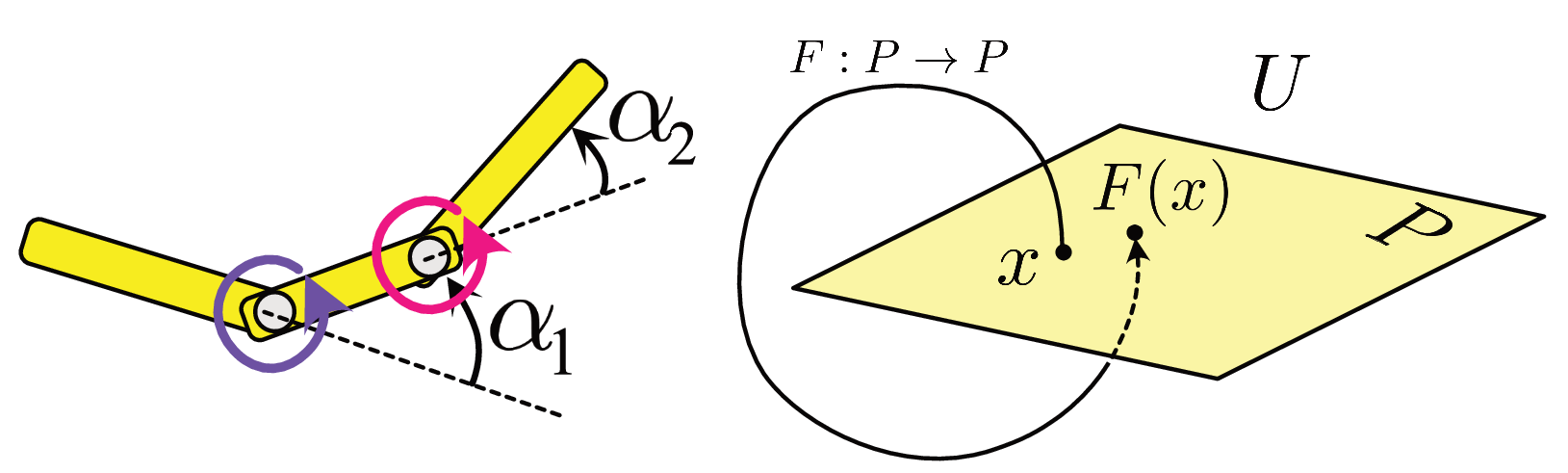}
\put(0,30){(a)}
\put(44,30){(b)}
\end{overpic}
\caption{(a) Schematic of Purcell's three-link model. (b) Schematic of Poincar\'e section $P$ in the state space $U$ and the associated Poincar\'e map $F:P\rightarrow P$. }
\label{fig:purcell}
\end{center}
\end{figure}

We first consider the case of $N=3$, where the three links are connected by two hinges [Fig. \ref{fig:purcell}(a)]. This is Purcell's three-link model \cite{purcell1977life}, which has been widely used as a simple and canonical mathematical model of microscale swimming \cite{hatton2013geometric, ishimoto2022self}. The state space $U$ is then four-dimensional and the Poincar\'e section $P$ is three-dimensional.

Before proceeding to the numerical results of the periodic orbits, we first analyse the condition where the periodic orbits exist. 
When $N=3$, the phase dynamics are written as
 \begin{equation}
 \left\{\begin{array}{ll}
 \dot{\phi}_1&=\omega_0+C \sin(\phi_2-\phi_1)+\sigma \cos(\phi_1) S_1(t) \\
 \dot {\phi}_2&=\omega_0+C \sin(\phi_1-\phi_2)+\sigma \cos(\phi_2) S_2(t) 
 \end{array}\right.
 \label{eq:P01},
 \end{equation}
and by introducing the phase difference $\Delta \phi = \phi_2-\phi_1$, we obtain the phase equation:
\begin{equation}
\Delta \dot \phi=-2C \sin(\Delta \phi)+\sigma \left[ \cos(\phi_2)S_2-\cos(\phi_1)S_1\right]
\label{eq:P02}.
\end{equation}
In the absence of mechanosensory feedback, namely, $\sigma= 0$, a complete in-phase synchronisation, $\Delta \phi=2n\pi$ ($n=0, \pm 1,\pm 2,\dots$), is a stable fixed point, while a complete anti-phase synchronisation, $\Delta \phi=(2n-1)\pi$, is an unstable fixed point.  

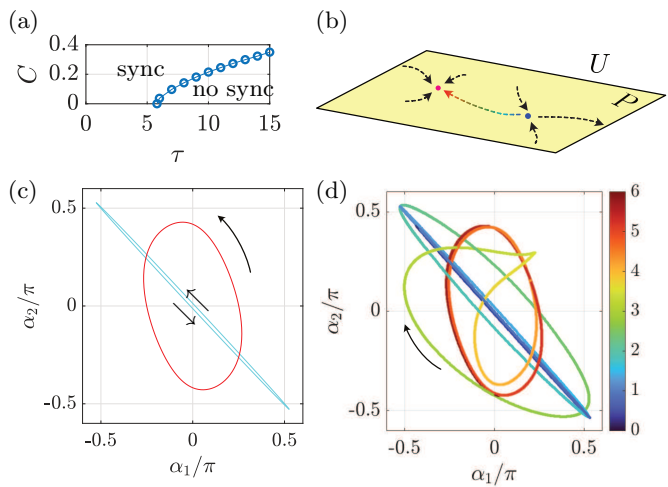
\begin{figure}[!t]
\begin{center}
\begin{overpic}[width=9cm]{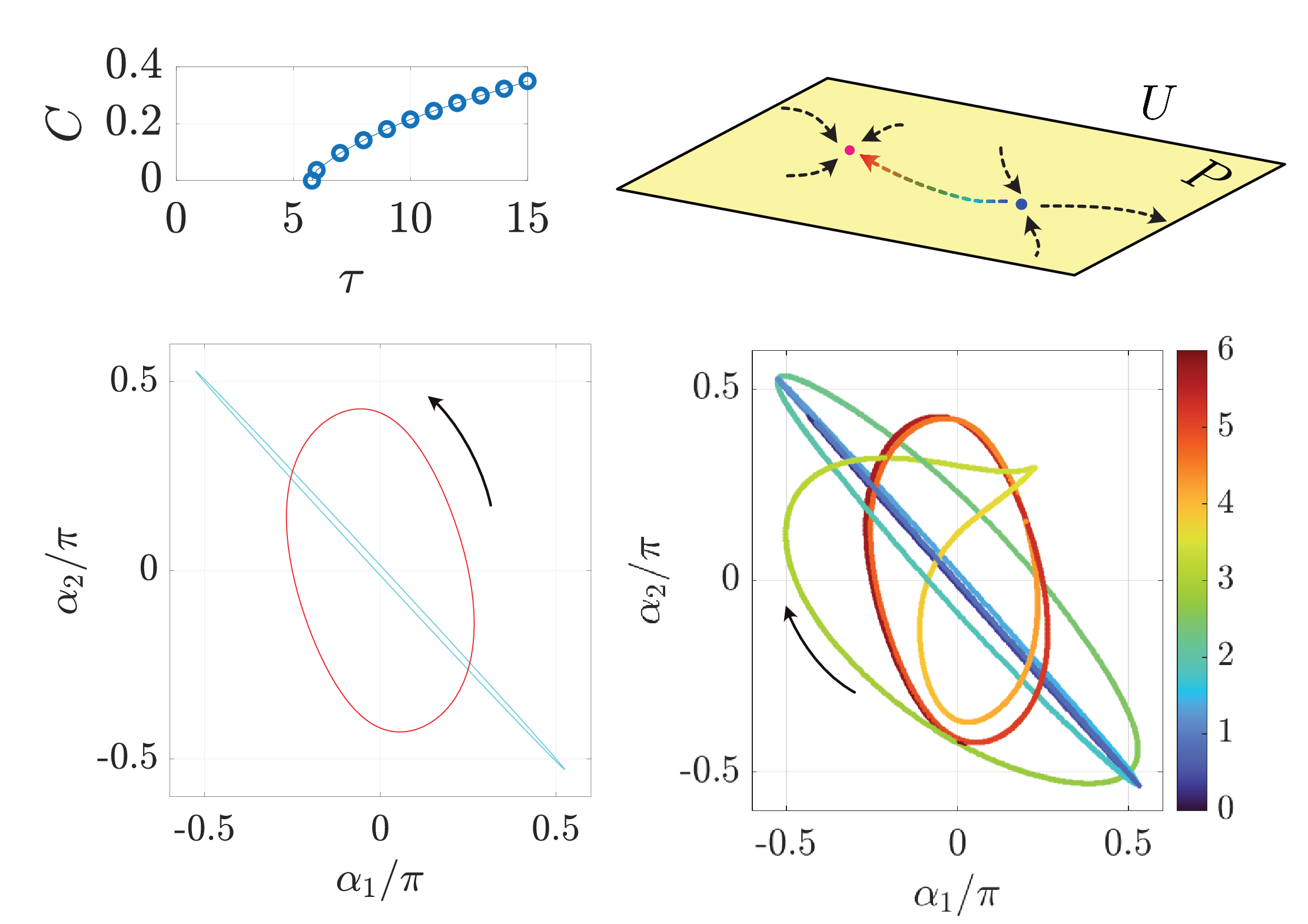}
\put(2,68){(a)}
\put(47,68){(b)}
\put(2,43){(c)}
\put(47,43){(d)}
\put(18,61){sync}
\put(29,58){no sync}
\put(26,25) {$\searrow$}
\put(28,27) {$\nwarrow$}
\end{overpic}
\caption{(a) Plots of critical value $C^\ast$ for different $\tau$ values for Purcell's swimmer in a low-viscosity medium. (b) Schematic of the dynamical systems on Poincar\'e section $P$. There is one stable fixed point (red) and one saddle point (blue). (c) Periodic orbits of the three-link swimmers in the $\alpha_1-\alpha_2$ shape space. The large orbit (red) is stable,  and the thinner orbit (blue) is unstable. Arrows indicate the direction of the trajectories. (d) Time evolution in the $\alpha_1-\alpha_2$ shape space, with the initial shape being set close to the unstable saddle. Different colours illustrate the time evolution.}
\label{fig:poincare}
\end{center}
\end{figure}

When the mechanosensory feedback is activated   ($\sigma\neq 0$), two synchronised states can still be observed,
which correspond to the in-phase and anti-phase synchronised states when $\sigma=0$, and they inherit their respective stability properties. 
In the small-amplitude case ($\tau\ll 1$),
we may analytically prove that these two synchronised states always exist, regardless of the values of $\textrm{Sp}, \gamma, \tau, \sigma$ and $C$.
The detailed derivations are provided in App. \ref{app:sync}. 

In the finite-amplitude case, however, the synchronised states can cease due to a saddle-node bifurcation. Indeed, there may be a critical coupling constant $C^{\ast}$, below which the synchronised states disappear. Fig. \ref{fig:poincare}(a) plots the values of the critical coupling constant for different sizes of $\tau$ with the other parameters being the same as in Fig. \ref{fig:sims}(a). At a large amplitude with $\tau\gtrsim 5.8$, phase slip occurs in the parameter region without synchronised states.

Nonetheless, this weak coupling region is very limited. In a large parameter region, which specifically contains the typical parameters for biologically relevant motion, there is one stable periodic orbit (SPO) and one unstable periodic orbit (UPO). We established through numerical stability analysis that the UPO is a saddle point. Fig. \ref{fig:poincare}(b) shows a schematic of the structure of the dynamical systems in $U$ and on $P$, where the SPO (the red point) and UPO (the blue point) are connected by heteroclinic orbits. By setting the parameter values as of the low-viscosity swimmer in Fig. \ref{fig:sims}(a), we numerically detect the  SPO and UPO. Fig. \ref{fig:poincare}(c) plots them in the $\alpha_1-\alpha_2$ shape space, with red and blue orbits representing the SPO and UPO, respectively. When the initial shape is located near the UPO, the dynamics evolve along a heteroclinic orbit, as shown in Fig. \ref{fig:poincare}(d). As time progresses (from blue to red), the trajectory departs from the UPO and finally approaches the SPO with the cyclic direction changing from anticlockwise to clockwise, to anticlockwise in the $\alpha_1-\alpha_2$ shape space.

\subsection{Three-link to $N$-link}


We now proceed to the analysis of the generic $N$-link system with $N>3$. Again, by the Poincar\'e map method, we numerically analysed the stability of the periodic orbits and found that only a single global SPO exists in a large parameter range in different media. 
Here, we show a typical behaviour of the model, by presenting the case of $N=10$ and crawling dynamics with the same parameters as in Fig. \ref{fig:sims}(c).

\begin{figure}[!t]
\vspace*{-0.5cm}
\begin{center}
\begin{overpic}[width=9cm]{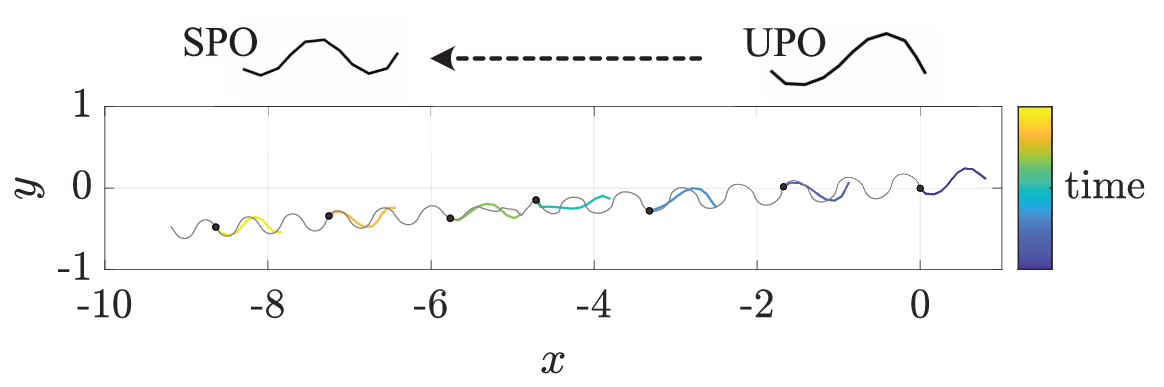}
\put(0,27){(a)}
\end{overpic}
\vspace{0.1cm}
\begin{overpic}[width=4.2cm]{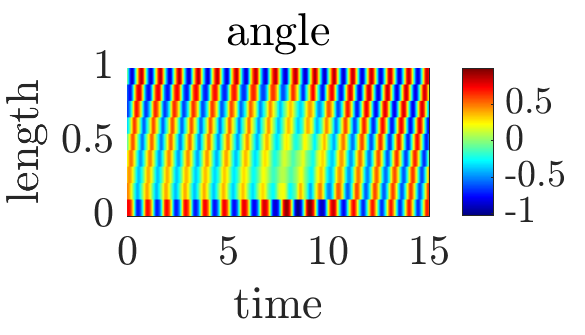}
\put(0,49){(b)}
\end{overpic}
\begin{overpic}[width=4.2cm]{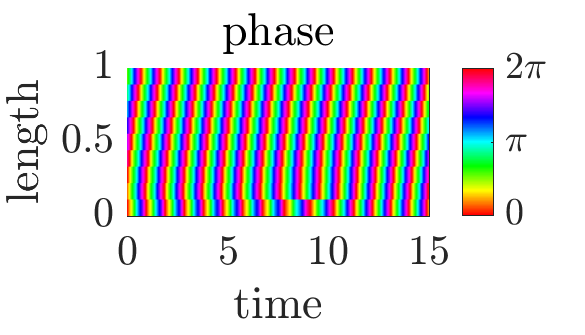}
\put(0,49){(c)}
\end{overpic}
\caption{Transient crawling locomotion from an unstable saddle (UPO) to a stable limit cycle (SPO) with $N=10$. (a) Snapshots of the shape and trajectory in the $xy$ plane. The object, initially located near the UPO, exhibits an eventual convergence to the SPO after transient behaviour. (b, c) Kymographs of the angle and phase dynamics, respectively.}
\label{fig:trans}
\end{center}
\end{figure}

With the same numerical analysis of the Poincar\'e section with the $N=10$ case, we detected a global SPO, by checking that all the Floquet exponents have a negative real part. A snapshot of the shape gait is shown in the top left of Fig. \ref{fig:trans}(a).
In the same parameter regions, we also found a UPO. As is shown in the top right of Fig. \ref{fig:trans}(a), the snapshot for this UPO shows a longer wavelength than that for the SPO waveform. We confirm that one of the Floquet exponents is a positive real number, indicating that the UPO is a saddle, like in the Purcell three-link model.

Hence, if the model configuration is initially close to the UPO, the shape and phase dynamics evolve along the heteroclinic orbits that connect the SPO and UPO, as shown in Fig. \ref{fig:trans}. The waveform shifts from the unstable longer wave to the stable shorter wave.

The global UPO becomes more robust as the number of links $N$ increases. In the parameter set of Fig. \ref{fig:trans}, the critical coupling constant, below which synchronisation ceases, is observed to be very small, at $C^\ast/\sigma\approx 0.01$.
Further, we found that large-scale stable locomotion emerges even for a larger $N$ ($\approx 100$), suggesting that only one SPO exists even for a larger $N$. 
In this section, we have seen a global attractor in the behavioural state space, together with transient behaviour. In the next section, we will use this structural property to manipulate the locomotor dynamics.


\section{Manoeuvring by modulating sensitivity}
\label{sec:manoeuver}

In the previous sections, we showed how our CPG-based model can achieve a stable locomotion pattern, with the mechanical feedback term playing a key role in the emergence of travelling waves. The strength of this term is measured by the sensitivity strength $\sigma$, which was observed to influence the gait aspect as well as the direction of motion depending on its sign. 
Hence, time variations of the sensitivity function are likely to allow changes in locomotion, such as a direction reversal or turn. 

In this section, we explore the behaviours obtained when following this idea and considering $\sigma$ as a function of time instead of a fixed parameter. The dynamical system governing the animal's motion now appears as a non-autonomous system, and its trajectory can be guided by tuning  $\sigma$ as a control function.

\subsection{Reverse motion}

\begin{figure}[!t]
\vspace*{0.1cm}
\begin{center}
~~~\begin{overpic}[width=8.4cm]{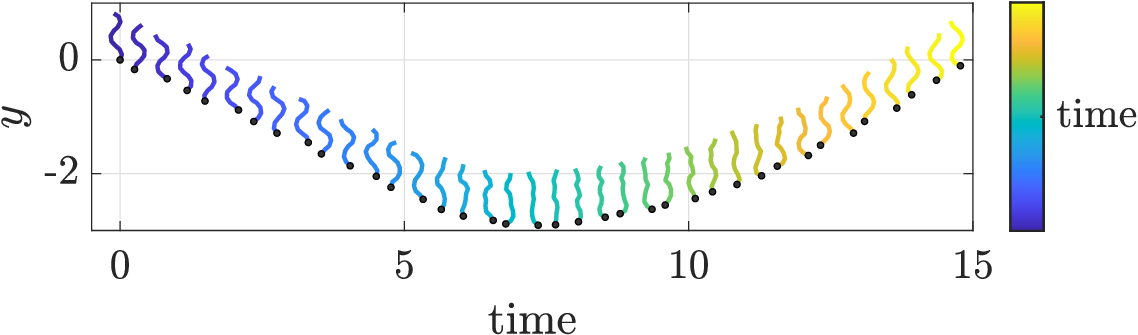}
\put(-1,29){(a)}
\put(25,23){$\big\downarrow$}
\put(70,23){$\big\uparrow$}
\end{overpic}
\vspace{0.1cm}
\begin{overpic}[width=4.2cm]{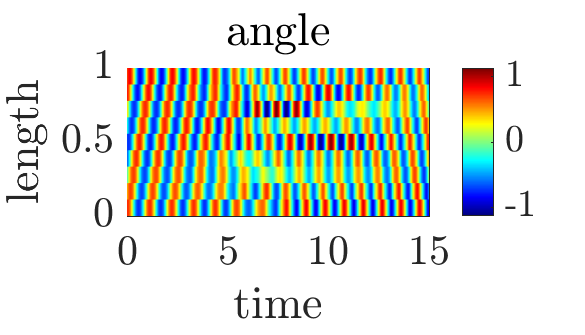}
\put(0,49){(b)}
\end{overpic}
\begin{overpic}[width=4.2cm]{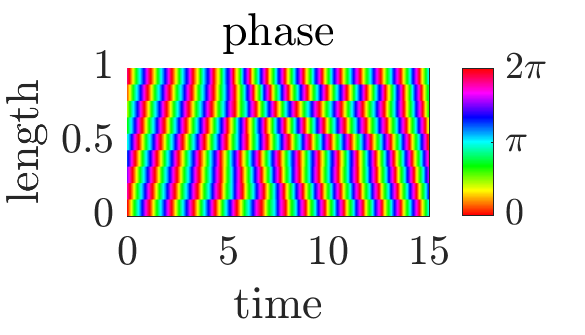}
\put(0,49){(c)}
\end{overpic}
\caption{Transient crawling behaviour when the sign of signal strength $\sigma$ switches. The SPO in Fig. \ref{fig:trans} was used for the initial condition. We used the same parameter set except for $\sigma$, which was switched to $-\sigma$ at time $t_0=0$. (a) Time series of the snapshots. The arrow indicates the movement direction, and the crawler moves in the $-y$ direction until it turns to reverse to a backward motion in the $+y$ direction.  (b, c) Kymographs of the angle and phase dynamics, respectively.}
\label{fig:switch}
\end{center}
\end{figure}

Sudden changes of the sensitivity value $\sigma$ with time were reported in a neuromechanical model of {\it C. elegans} crawling motion as a plausible cause for motion reversal and turns \cite{olivares2018potential}. 

From the point of view of the dynamical system, Eq. \eqref{eq:M02}, a simple sign change of $\sigma$ at some time $t_0$ modifies the dynamics landscape and in particular the position of the global SPO, which we know robustly emerges for both $\sigma$ and $-\sigma$. Thus, forward-backward motion reversal naturally occurs upon this sensitivity change.

When the coupling strength $C$ is sufficiently large, the locomotion immediately converges to the global SPO of the new dynamical system, and the crawler exhibits a quick reverse motion, which agrees well with {\it C. elegans} observations \cite{stephens2008dimensionality}. 

When the coupling weakens, the transient dynamics between the motion reversal exhibit more complicated behaviours, as shown in Fig. \ref{fig:switch}, where we present the transient behaviours of crawling locomotion with the same parameter set as in Fig. \ref{fig:trans}. We employed the SPO of Fig. \ref{fig:trans} as the initial configuration and changed the sign of $\sigma$ as $\sigma=12\mapsto\sigma=-12$ at time $t_0=0$. Snapshots of the shape gait are shown, with the horizontal axis indicating the time evolution. The arrow indicates the movement direction, and the crawler moves in the $-y$ direction until it turns to a backward motion moving in the $+y$ direction with a transient regime without phase synchrony. As plotted in Fig. \ref{fig:switch}(b,c), the kymographs of the angle and phase dynamics also show complex transient dynamics until the angle reformulates the backward travelling wave. 


\subsection{Turning behaviour}
\label{sec:turn}

So far, we have observed forward and backward motion, as well as the transition between both when the sensitivity strength $\sigma$ is switched between a set value and its opposite. 
However, this abrupt switch can be modulated to allow manoeuvring and reaction to changes in the environment. This would particularly enable foraging strategies such as run-and-tumble or any kind of taxis. 

To illustrate this, we take the omega-turn behaviour of {\it C. elegans} as an example and propose a way to reproduce it in our CPG model.
The turning strategy of {\it C. elegans} on agar gels is called an 'omega-turn' because it temporarily breaks the symmetry of its gait by curving towards one side in a shape that roughly mimics the Greek letter $\Omega$ \cite{stephens2008dimensionality, pierce2008genetic}. 

To allow these occasional symmetry breaks within the CPG dynamics, we add a term in Eq. \eqref{eq:M02} governing the evolution of $\phi$:
\begin{multline}
    \dot{\phi}_i = \omega_0 + \sum_j C \sin (\phi_j - \phi_i) + \sigma_1(t) \cos(\phi_i) S_i(t) \\ 
    + \sigma_2(t) K \cos(\phi_i - \psi_i),
    \label{eq:omega_1}
\end{multline}
where $K$ and $\psi_i$ are set to induce the local angle $\alpha_i$ to curve into some angle $\alpha_i^{\Omega}$ and the amplitude of $\sigma_2$ measures this forcing strength. Details on how to design $K$ and $\psi_i$ are provided in App. \ref{sec:maneouvering functions}.

\begin{figure*}
    \centering
    \begin{overpic}[height=0.352\textwidth]{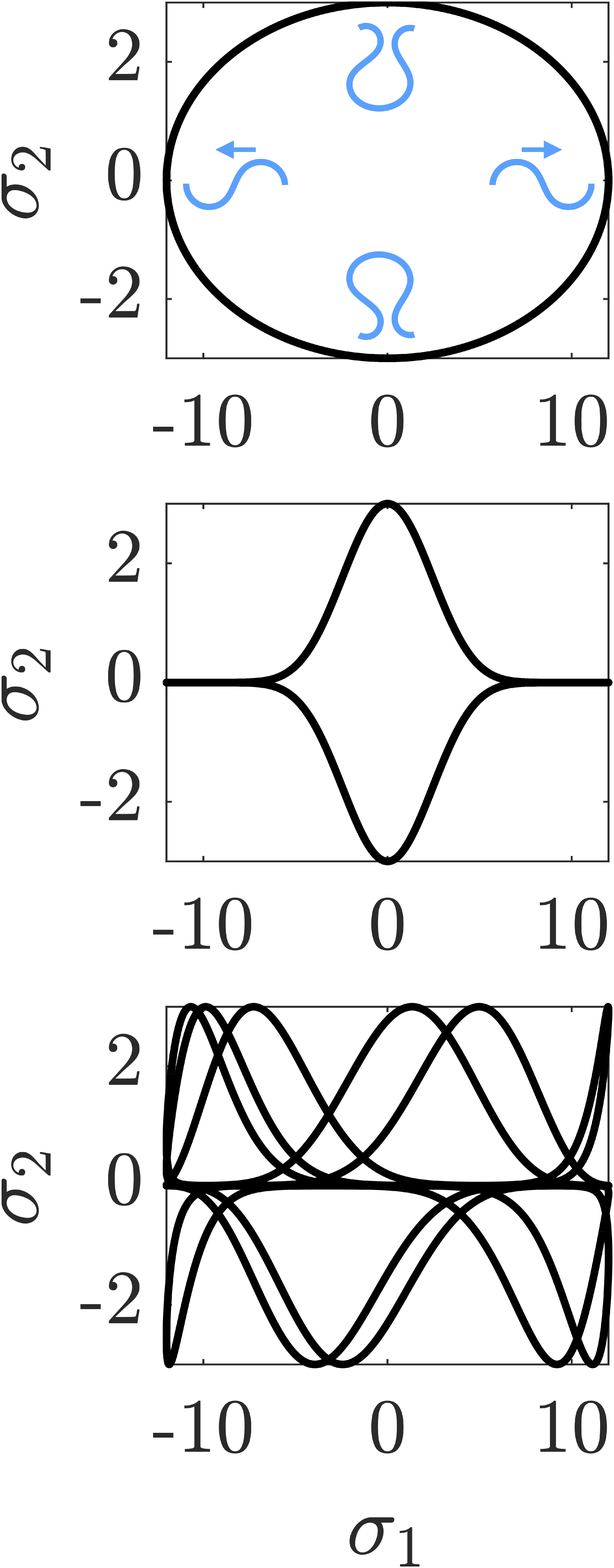}
    \put(0,98){(a)}
    \end{overpic}
    \begin{overpic}[height=0.352\textwidth]{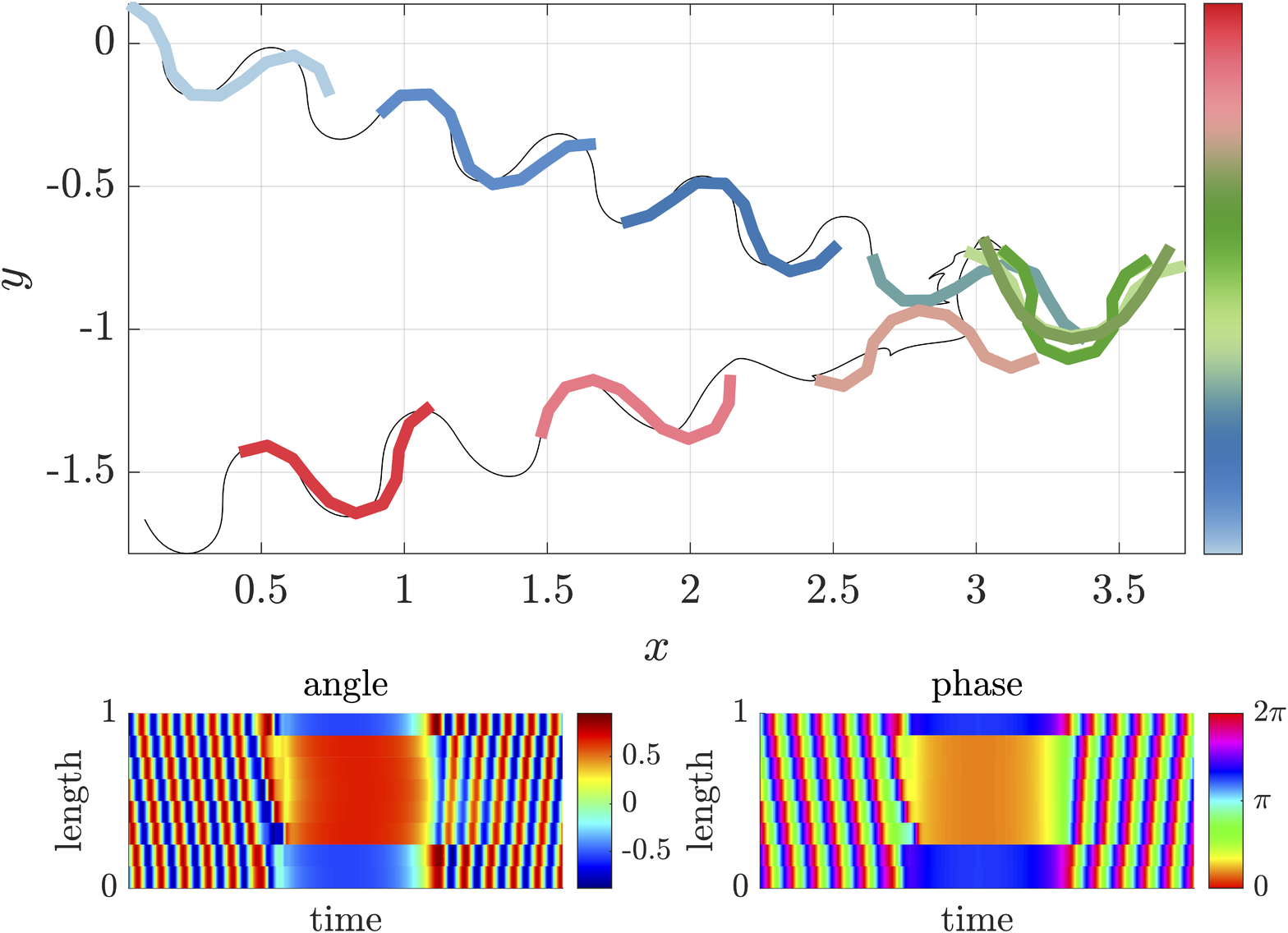}
    \put(0,77){(b)}
    \end{overpic}
    \begin{overpic}[width=0.3564\textwidth]{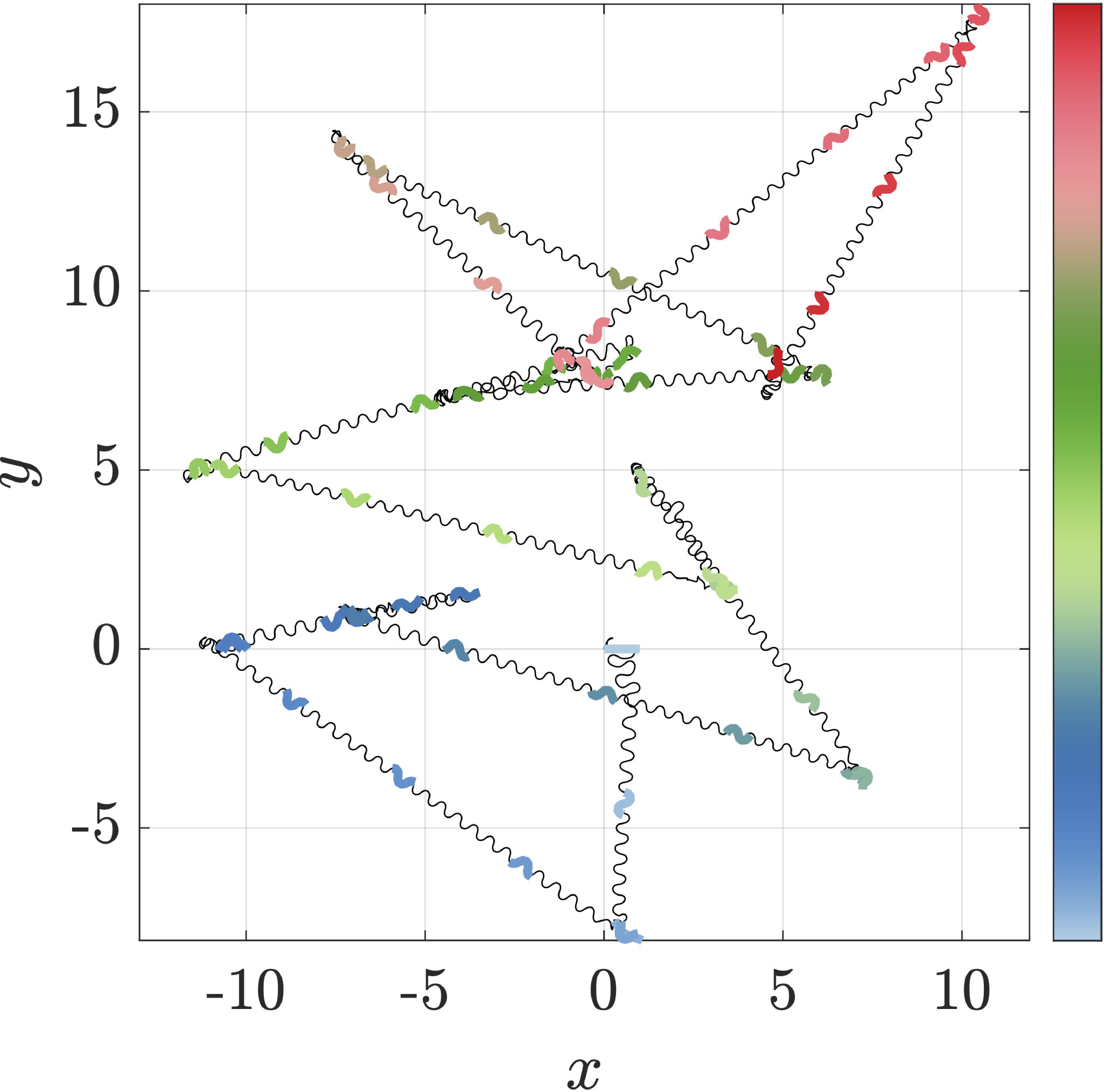}
    \put(0,98){(c)}
    \end{overpic}
    \caption{Simulated turning and foraging behaviour. (a) Representation of the control functions $\sigma_1(t),\sigma_2(t)$ in the $(\sigma_1,\sigma_2)$ plane. The top panel summarises the different states preferred by the model animal at each extremity of the activation pattern: (clockwise from the top) omega-turn, forward motion, mirrored omega-turn and backward motion. The middle and bottom panels respectively display the activation functions used to generate the trajectories of panels (b) and (c). (b) Detailed aspect of the omega-turn simulation, with associated angle and phase kymographs. (c) Long-time behaviour with decorrelated frequencies for $\sigma_1$ and $\sigma_2$, showing complex, irregular trajectories.}
    \label{fig:omega}
\end{figure*}

To clarify the role of the two activation functions in Eq. \eqref{eq:omega_1}, the factor term of $\sigma_1$ sustains periodic locomotion as discussed previously, while the factor term in $\sigma_2$ drives each $\phi_i$ to converge to a constant value in a way that immobilises the model animal in a given shape measured by target angles $\alpha_i^\Omega$. The relative amplitude of $\sigma_1$ and $\sigma_2$ determines which of these two behaviours dominate the instantaneous dynamics.

With this setup, when tuning $\sigma_1$ and $\sigma_2$ with respect to time, for example following periodic functions, we may observe alternating forward or backward motion and also omega-turn-like behaviours effectively reorienting the animal's direction of motion between forward and backward sections of the trajectory [top diagram of Fig. \ref{fig:omega}(a)], where we used the same parameter set of the crawling locomotion as in previous figures.

A family of functions for $\sigma_1$ and $\sigma_2$ that may yield an interesting range of behaviours is given by
\begin{equation}
\left \{ \begin{array}{l}
\displaystyle \sigma_1(t) = \sigma_1^0 \cos^{2n_1+1}\left ( \frac{t}{T_1} + \chi \right ),\\
\displaystyle \sigma_2(t) = \sigma_2^0 \sin^{2n_2+1}\left ( \frac{t}{T_2} \right ).
\end{array}
\right.
\label{eq: sigma}
\end{equation}
The integers $n_1$ and $n_2$ should be relatively small to obtain the distributed activation, while large values simulate impulse-like behaviour. This latter option is suitable for the omega-turn activation. The corresponding activation function is plotted in the $\sigma_1-\sigma_2$ plane in the middle panel of Fig. \ref{fig:omega}(a), and the resulting short-term trajectory is shown in Fig. \ref{fig:omega}(b). The snapshots of the shape gait are illustrated in different colours from blue to green to red as time progresses. The model animal initially moves towards the right, but them briefly turns and then starts moving again towards the left.  

This mechanism can be exploited to simulate the animal exploring its environment. Further variability may be obtained by desynchronising the periodicity of $\sigma_1$ and $\sigma_2$, taking for example, $T_2/T_1 = \sqrt{2}$ in Eq. \eqref{eq: sigma}; the corresponding activation function is shown in the bottom plot of Fig. \ref{fig:omega}(a), and the resulting long-term trajectory is displayed in Fig. \ref{fig:omega}(c). Sharp turns at seemingly unpredictable angles due to the combination of the omega-turn mechanism and complex activation functions makes the trajectory strongly reminiscent of foraging behaviour \cite{stephens2008dimensionality}.

\section{Concluding remarks}
\label{sec:conc}

In this study, we formulated a minimal model of neuromechanical undulatory locomotion of an elastic slender body in a dissipation-dominated environment using the resistive force theory and coupled phase oscillators that form a CPG-like network.
In contrast to existing CPG models, which assume an undulatory body motion, our network model preserves the motion reciprocity. 
Nonetheless, local phase modulation of each oscillator gives rise to robust 
undulatory locomotion in a self-organised manner, through mechanosensory feedback of the local mechanical load.


By numerical explorations, we obtained a stable, coherent undulatory motion in a broad parameter region and the emergent gait pattern reasonably agrees with various motions of {\it C.elegant} in different media. This process emerges from local symmetry breaking in the fore-aft resistive torque perceived by mechanoreceptors, which is then reflected at the network level by symmetry breaking with respect to the swimmer's transverse plane, allowing non-reciprocal movement. This elementary mechanism thus explains the link between local and global dynamics at the origin of the motor activation wave, a property that could be extended to other sensory feedback, such as proprioception.

This coherent undulatory locomotion was further explored using Purcell's three-link model. We found through theoretical and numerical analyses that a saddle-node bifurcation of periodic orbits occurs at a certain level of phase coupling. 
This indicates that the local CPG oscillators require a certain level of coupling to stabilise the network and that the sensory feedback cannot generate a stable gait on its own.
The global stable periodic orbit is more robust when increasing the number of links, $N$, and we found the basic structure of the dynamical system to be similar to that of the Purcell model. Nonetheless, fine structures, such as the presence and number of UPOs and the corresponding trajectories connecting them to the main SPO, have not been fully analysed because of the large degrees of freedom, which are left for future work.

We then exploited this dynamical system structure to design manoeuvring strategies, by using the signal sensitivity as a control. Motivated by {\it C. elegans} behaviour on agar, we demonstrated through numerical explorations that our CPG model well reproduces the reverse motion and foraging behaviour known as the omega-turn. Our deterministic model can thus include various predefined stereotyped behaviours, the so-called neuromechanical `templates' \cite{full1999templates} given by cycles and fixed points in the state space. Similarly, our model exhibits apparently random motion due to transient dynamics between these attractors. Recent data-driven modelling approaches may discover such attractors and the associated transients in the behavioural state space \cite{ahamed2021capturing, gilpin2020deep, valani2023attractor}

One can also build upon this property  to investigate, with a set of simple building blocks, complex long-term behaviours observed in some biological locomotors. Furthermore, this methodology of manipulating the nonlinear dynamical systems might offer a robust design method for redundant robot controllers \cite{seeja2022survey}. 


In conclusion, our dynamical systems viewpoint is useful not only for understanding the extent of complexity and diversity found in animal locomotion but also for designing self-propelled, self-adaptive robots for a wide range of dissipative environments.

\section*{Acknowledgments}
K.I. acknowledges the Japan Society for the Promotion of Science (JSPS) KAKENHI for Transformative Research Areas A (Grant No. 21H05309) and the Japan Science and Technology Agency (JST), FOREST (Grant No. JPMJFR212N). C.M. acknowledges funding support for JSPS International Research Fellow (PE22023) (Grant No. 22KF0197). J.H was supported by the French National Research Agency ANR (Grant No. ANR-2019-CE33-0004-01). K.I. and C.M. were supported in part by the Research Institute for Mathematical Sciences, an International Joint Usage/Research Center located at Kyoto University. The authors acknowledge Prof. Toshiyuki Nakagaki for fruitful discussion.

\begin{appendix}


 \section{Phase synchronisation of the three-link model}
\label{app:sync}
 
In this appendix, we consider the condition for the existence of phase synchronisation in a small-amplitude regime, by analysing the nonlinear term from the mechanosensory feedback in Eq.\eqref{eq:P02}.

\subsection{Asymptotic slow dynamics of the coupled-phase oscillators}

The sensory signals, $S_1$ and $S_2$, are generally nonlinear in the phase variables, as these signals result from body-environment interactions. Nonetheless, it is reasonable to assume that after a transient the system reaches periodic dynamics, then the sensory signals $S_1$ and $S_2$ can be expanded by a Fourier series of $\phi_1$ and $\phi_2$ as
 \begin{equation}
 \left\{\begin{array}{ll}
 S_1&=\hat S_{1,1} \cos( \phi_1+\psi_{1,1}) +\hat S_{1,2} \cos( \phi_2+\psi_{1,2})+\dots,  \\ \\
 S_2&=\hat S_{2,1} \cos( \phi_1+\psi_{2,1}) +\hat S_{2,2} \cos( \phi_2+\psi_{2,2})+\dots, 
 \end{array}\right.
 \label{eq:P04}
 \end{equation}  
where the coefficients $\{\hat S_{I,j} \}$ and $\{\psi_{i,j} \}$ are interpreted as the gain and delay, respectively, due to motor activation at each hinge. In Eq. \eqref{eq:P04}, we only expressed the modes that oscillate in the same frequency as $\phi_1$ and $\phi_2$, because the other Fourier modes do not affect the slow dynamics, which we analyse below.


To obtain the evolution of the phase difference dynamics, we substitute the expansions \eqref{eq:P04} into Eq. \eqref{eq:P02}, and then drop the trigonometric terms including $\phi_1+\phi_2$, $2\phi_1$ and $2\phi_2$, which oscillate with the higher frequency of $2\omega_0 t$. The resulting equation is
 \begin{equation}
 \begin{array}{ll}
\Delta \dot \phi=& \displaystyle{-2C \sin(\Delta \phi)+\frac{\sigma}{2} \left[\hat S_{2,2} \cos(\psi_{2,2})  -\hat S_{1,1} \cos(\psi_{1,1})\right]} \\ \\& \displaystyle{+\frac{\sigma}{2} \left[\hat S_{2,1}\cos(\Delta \phi-\psi_{2,1})
-\hat S_{1,2}\cos(\Delta \phi+\psi_{1,2}) \right]}.
 \end{array}
 \label{eq:P04b}
 \end{equation}
We can further simplify this by introducing the frequency shift $\Delta \omega$,
 \begin{equation}
 \Delta \omega=\frac{1}{2} \left[\hat S_{2,2} \cos(\psi_{2,2})  -\hat S_{1,1} \cos(\psi_{1,1})\right]
 \label{eq:P05},
 \end{equation}
 and two parameters, $K$ and $\Psi$, which correspond to the amplitude and delay, respectively,
 to obtain
 \begin{equation} 
\Delta \dot \phi=  \sigma \Delta \omega+K \sin(\Delta \phi-\Psi)
\label{eq:P06}.
 \end{equation} 
 
The necessary condition for synchronisation is thus found as
\begin{equation}
K> \sigma \vert \Delta \omega \vert
\label{eq:P07}.
\end{equation}
Otherwise, $\Delta \phi$ is not bounded. If this condition is satisfied, there are two steady solutions, and
the stability of these solutions follows from the sign of $K \cos \left( \Delta \phi \right)$.

 \subsection{Small-amplitude motion}

We now proceed to examining the condition \eqref{eq:P07} under a small amplitude, specifically, $|\alpha_1|,|\alpha_2|\ll 1$.

The resulting phase dynamics are used to activate the torque on each hinge. Because the bending torque is a sub-leading contribution under the small-amplitude condition, the torque expression is simply given by
 \begin{equation}
 \tau_1=\tau \cos( \phi_1),~
 \tau_2=\tau \cos( \phi_2)  
 \label{eq:P03}.
 \end{equation}  

Our feedback signal, the local torque load, $S_i$, is a linear function of velocity $\dot{\bm{z}}$ in a dissipation-dominated environment. These values are obtained by the linear problem, Eq. \eqref{eq:M03}, and the generalised resistance matrix ${\bf A}(\bm{\alpha})$ is evaluated as the value at $\bm{\alpha}=\bm{0}$ under the small-amplitude condition. Due to this linear property, the sensory signals are proportional to $\cos\phi_1$ and $\cos\phi_2$, and this fact requires that all the delay factors in Eq. \eqref{eq:P03} should vanish as $\psi_{1,1}=\psi_{1,2}=\psi_{2,1}=\psi_{2,2}=0$.
The explicit form of the parameters in \eqref{eq:P06} are then given by
\begin{eqnarray}
\Delta\omega&=&\frac{1}{2}\left(\hat{S}_{2,2}-\hat{S}_{1,1}\right)
\label{eq:10a},\\
K&=&\left[\frac{\sigma^2}{4}\left(\hat{S}_{2,1}-\hat{S}_{1,2}\right)^2+4C^2 \right]^{\frac{1}{2}}
\label{eq:10b},\\
\Psi&=&\arctan\left[\frac{\sigma}{4C}\left(\hat{S}_{2,1}-\hat{S}_{1,2}\right)\right]
\label{eq:10c},
\end{eqnarray}
and condition \eqref{eq:P07} is reduced to
\begin{equation}
\left(\frac{C}{\sigma}\right)^2>\frac{1}{16}\left[\left(\hat{S}_{2,2}-\hat{S}_{1,1}\right)^2-\left(\hat{S}_{2,1}-\hat{S}_{1,2}\right)^2\right]
\label{eq:P11}.
\end{equation}

An explicit form of ${\bf A}(\alpha)$ is given in the Appendix of \cite{moreau2019local}, and the torque on each link can also be calculated from RFT. By performing the matrix calculations above, we can explicitly compute the entries of $\{\hat{S}_{i,j}\}$ in the dimensionless form as
\begin{equation}
{\bf \hat{S}}=
    \begin{pmatrix}
       \hat{S}_{1,1} & \hat{S}_{1,2}\\
       \hat{S}_{2,1} & \hat{S}_{2,2}
    \end{pmatrix}
   =\frac{\tau}{5}\begin{pmatrix}
       -4 & -6 \\ -34 & 14
   \end{pmatrix} 
\label{eq:P12}.
\end{equation}
These values are purely geometrical and do not depend on mechanical parameters such as $c_{\perp}$ and $\kappa$. It is readily found that the right-hand side of Eq. \eqref{eq:P11} is negative; hence, the synchronised steady states always exist with one stable and one unstable fixed point in the phase difference.  

\section{Manoeuvring functions}
\label{sec:maneouvering functions}

In this appendix, we detail how to design the additional term in Sec. \ref{sec:turn}, allowing us to reproduce a turning strategy inspired by {\it C. elegans}'s omega-turn behaviour.  
 
 Recall that we aim at driving the phase dynamics such that the body is immobilised in some preset $\Omega$ shape described by the angles $\alpha_i^{\Omega}$.
Because the torque on junction $i$ is given by 
\begin{equation}
\tau_i = \tau \cos \phi_i - \kappa \alpha_i,
\end{equation} 
it is clear that having $\phi = \phi_i^{\Omega}$, with
\begin{equation}
    \phi_i^{\Omega} = \cos^{-1} \left ( \frac{\kappa \alpha_i^{\Omega}}{\tau} \right )
    \label{eq:set_point_phase},
\end{equation}
would induce
\begin{equation}
    \tau_i = -\kappa (\alpha_i - \alpha_i^{\Omega}),
\end{equation}
effectively setting $\alpha_i^{\Omega}$ as a stable equilibrium point for the angle dynamics. Note that this forbids giving $\alpha_i^{\Omega}\geqslant \tau/\kappa$ a sufficient torque amplitude for a given stiffness $\kappa$ to reach the set point $\alpha_i^{\Omega}$.


As a first approach, we consider an oscillator with an intrinsic frequency $\omega_0$ subjected to an external signal of constant amplitude $\sigma_2$ and local phase $\psi_i$,
\begin{equation}
    \dot{\phi}_i = \omega_0 - \sigma_2 \cos{(\phi_i -\psi_i)},
    \label{eq:phi dynamics}
\end{equation}
and aim to lock the phase $\phi_i$ to the reference phase $\phi_i^{\Omega}$ via the parameters $(\sigma_2,\psi_i)$.
Straightforward analysis of Eq. \eqref{eq:phi dynamics} shows that this phase equation has a stable fixed point given by 
\begin{equation}
    \phi_i^0 = 2 \pi -   2 \cos^{-1} \left ( \frac{\omega_0}{\sigma_2} \right ) - \psi_i,
    \label{eq:phase_locking}
\end{equation}

\noindent for ${\sigma_2}>\omega_0$. The parameter $\sigma_2$ is set to control the convergence time to the target angle. By combining   Eqs. \eqref{eq:set_point_phase} and \eqref{eq:phase_locking},   the   $\Omega$ shape is a stable fixed point if   the phases $\psi_i$     satisfies 
\begin{equation}
    \psi_i = 2 \pi - \cos^{-1} \left ( \frac{\kappa \alpha_i^{\Omega}}{\tau} \right )  - \cos^{-1} \left ( \frac{\omega_0}{ \sigma_2}  \right ) .
\end{equation}

Hence, the manoeuvring function takes $\alpha_i^{\Omega}$ as input and $\sigma_2$ as a free parameter to tune the attractivity of the fixed point. 



In the full-phase dynamics given by Eq. \eqref{eq:omega_1}, this phase-locking dynamics competes against the CPG term and sensory feedback. Dimensional analysis suggests that phase locking occurs when the ratio $C/\sigma_2$ and $\sigma_1/\sigma_2$ are significantly small with respect to $\omega/\sigma_2<1$, which is consistent with our observations.  

In Sec. \ref{sec:manoeuver}, we use $\sigma_2$ as a control function and tune it with respect to time to make it sporadically dominate the dynamics, alternatively with the locomotion term $\sigma_1$.

\end{appendix}
\bibliography{library}

\end{document}